\documentclass[,superscriptaddress,twocolumn,preprintnumbers,amssymb,nofootinbib]{revtex4-1}
\usepackage{amsmath, amsthm, amssymb} 

\usepackage{graphics,bm}
\usepackage{epsfig}
\usepackage{graphicx}
\usepackage{amsmath}
\usepackage{wrapfig}
\usepackage{color}
\usepackage{tikz}
\usetikzlibrary{arrows,shapes}
\usetikzlibrary{trees}
\usetikzlibrary{matrix,arrows} 			
\usetikzlibrary{positioning}				
\usetikzlibrary{calc,through}				
\usepackage{pgffor}							

\usetikzlibrary{decorations.pathmorphing}	
\usetikzlibrary{decorations.markings}
\tikzset{
    sigmaCT/.style={draw=black, postaction={decorate},
        decoration={markings,mark=at position .99 with {\arrow[draw=black]{>}},mark=at 		 position .99 with {\arrow[
draw=black]{<}}}},
    pionCT/.style={dashed,draw=black, postaction={decorate},
        decoration={markings,mark=at position .99 with {\arrow[draw=black]{>}},mark=at position .99 with {\arrow[draw=black]{<}}}},    
    fermionCT/.style={draw=black, postaction={decorate},
        decoration={markings,mark=at position .5 with {\arrow[draw=black]{>}},mark=at position .99 with {\arrow[draw=black]{>}},mark=at position .99 with {\arrow[draw=black]{<}}}},    
    fermion/.style={draw=black, postaction={decorate},
        decoration={markings,mark=at position .55 with {\arrow[draw=black]{>}}}},
    fermionbar/.style={draw=black, postaction={decorate},
        decoration={markings,mark=at position .55 with {\arrow[draw=black]{<}}}},
    pion/.style={dashed,draw=black, postaction={decorate}},
    sigma/.style={draw=black, postaction={decorate}}
}

\newcommand{\beq}{\begin{equation}}
\newcommand{\eeq}{\end{equation}}
\newcommand{\bqa}{\begin{eqnarray}}
\newcommand{\eqa}{\end{eqnarray}}

\newcommand{\os}{\text{\tiny OS}}
\newcommand{\ms}{\overline{\text{\tiny MS}}}
\newcommand{\bare}{\text{\tiny B}}

\def\sumint{\hbox{$\sum$}\!\!\!\!\!\!\int}
\def\square{\vcenter{\vbox{\hrule height.4pt
          \hbox{\vrule width.4pt height4pt
          \kern4pt\vrule width.3pt}\hrule height.4pt}}}

\begin{document}

\title{On-shell parameter fixing in the quark-meson model}
\author{Prabal Adhikari}
\email{adhika1@stolaf.edu}
\affiliation{St. Olaf College, Physics Department, 1520 St. Olaf Avenue,
Northfield, MN 55057, USA}

\author{Jens O. Andersen}
\email{andersen@tf.phys.ntnu.no}
\affiliation{Department of Physics, Faculty of Natural Sciences, NTNU, 
Norwegian University of Science and Technology, H{\o}gskoleringen 5,
N-7491 Trondheim, Norway}
\affiliation{Niels Bohr International Academy, 
Blegdamsvej 17, Copenhagen 2100, Denmark}
\author{Patrick Kneschke}
\email{patrick.kneschke@uis.no}
\affiliation{Faculty of Science and Technology, University of Stavanger,
N-4036 Stavanger, Norway}
\date{\today}

\begin{abstract}
The quark-meson model is often used as an effective low-energy
model for QCD to study the chiral transition at finite temperature $T$
and baryon chemical potential $\mu_B$.
The parameters in the quark-meson model can be found by expressing
them in terms of the sigma mass $m_{\sigma}$, the pion mass $m_{\pi}$, 
the constituent quark mass $m_q$ and the 
pion decay constant $f_{\pi}$. In practice, this matching is done at tree level,
which is inconsistent once we take loop effects of the effective potential
into account. We show how to properly perform the matching in the 
quark-meson model by using the on-shell and the modified
minimal subtraction renormalization schemes relating the physical
masses and the pion decay constant to the running mass parameter
and couplings. 
We map out the phase diagram in the $\mu_B$--$T$
plane and compare our results with other approximations.

\end{abstract}

\maketitle

\section{Introduction}
The first phase diagram of quantum chromodynamics (QCD) appeared
in the 1970s, where it was suggested that it consists of
a confined low-temperature phase of hadrons and a deconfined 
high-temperature phase of quarks and gluons.  
Since the appearance of this phase diagram, 
large efforts have been made to map it out in detail. 
The only existing first-principles method used to calculate the 
properties of finite-temperature below the chiral transition
is lattice QCD. However, due to the sign problem, it is difficult
to perform lattice simulations at finite baryon chemical potential.
Mapping out the phase diagram is therefore based on model calculations,
in particular in the region of low temperature and large baryon chemical
potential. See Refs. \cite{rev1,rev2} for reviews.

The $O(4)$-symmetric linear sigma model (LSM) is probably the simplest
low-energy model of QCD. The degrees of freedom are the pions and the
sigma particle. Often this model is augmented by an isospin doublet
of fermions.
In the old days, the fermionic doublet was identified with the neutron and
proton. Now the isospin doublet consists of a $u$ and a $d$ quark.
This extended model is referred to as the quark-meson (QM) model
or the linear sigma model with quarks (LSMq). One may object to having
both quark and mesonic degrees of freedom present at the same time.
At very low temperatures, this is a valid objection since quarks are confined.
This has led to the introduction of the Polyakov loop in these models
in order to mimic confinement in QCD in a statistical sense
by coupling the chiral models to a constant $SU(N_c)$ background
gauge field $A_{\mu}^a$ \cite{fuku}.
One can express this background gauge field in terms of the
complex-valued Polyakov loop variable $\Phi$ and consequently
the effective potential becomes a function of the
expectation value of the chiral condensate and 
the expectation value of the Polykov loop. The latter then serves
as an approximate order parameter for confinement \cite{benji}.
Finally, one adds the contribution to the free 
energy density from the gluons via a phenomenological Polyakov loop
potential \cite{ratti,fuku2}.
At finite temperature and chemical potential, the (P)QM is often
treated in the large-$N_c$ limit which implies that one takes into account
the one-loop correction to the effective potential from the fermions,
but treats the mesonic degrees of freedom at tree level \cite{scav}.
In some cases, one also neglects the vacuum fluctuations from 
the fermions and therefore renormalization issues altogether.
This is sometimes referred to as the ``no-sea'' approximation.

The Lagrangian of the QM model has several para\-meters that can be expressed
in terms of the physical quantities $m_{\sigma}$, $m_{\pi}$, $m_q$,
and $f_{\pi}$.
In this way one can fix the
parameters of the model such that it reproduces the vacuum physics
correctly. 
However, in most renormalization schemes, the tree-level relations between
the parameters in the Lagrangian and physical quantities receive
radiative corrections. It is therefore inconsistent to use tree-level
values for these parameters in for example the calculation 
of the effective potential.
While the on-shell parameters
take their tree-level values, the parameters in $\overline{\rm MS}$ scheme
are running and depend on the renormalization scale $\Lambda$, which
has been introduced to keep the canonical dimension of the loop integrals.
The idea is then to calculate the counterterms in the
on-shell scheme as well as in the $\overline{\rm MS}$ scheme and relate
the renormalized parameters in the two. 
The calculation of the effective potential is then carried out using
(modified) minimal subtraction and the relations between the running
parameters and the on-shell parameters, i.e. physical quantities
are then used as input.
This procedure has been well-known
for decades by people doing loop
calculations in the Standard Model, 
\cite{sirlin1,sirlin2,hollik,hollik2}, but seems
not to have been appreciated by practitioners in finite-temperature
field theory, see however Refs. \cite{laine,chiku0,chiku,hidaka,fix1}.

The paper is organized as follows.
In Sec. II we briefly discuss the quark-meson model.
We also calculate the self-energies and extract the counterterms
in the on-shell scheme.
In Sec. III, we
derive relations between the physical quantities and the running parameters.
In Sec. IV, we apply our results to the quark-meson model to
map out the phase diagram in the $\mu$--$T$ plane.
In the appendix, we list the integrals that are necessary in our 
calculations.

\section{Quark-meson model}
In this section we briefly discuss the quark-meson model and
calculate the one-loop self-energies in the large-$N_c$ limit.
We also derive the counterterms in the on-shell scheme.
\subsection{Lagrangian and self-energies}
The Lagrangian of the two-flavor quark-meson model 
in Minkowski space is 
\bqa\nonumber
{\cal L}&=&
{1\over2}\left[(\partial_{\mu}\sigma)^2
+(\partial_{\mu}{\boldsymbol \pi})^2\right]
-{1\over2}m^2(\sigma^2+{\boldsymbol\pi}^2)
\\&&
\nonumber
-{\lambda\over24}(\sigma^2+{\boldsymbol\pi}^2)^2
+h\sigma+\bar{\psi}
\left[
i/\!\!\!\partial
+(\mu+\mbox{$1\over2$}\tau_3\mu_I)\gamma^0
\right.\\ && \left.
-g
(\sigma+i\gamma^5{\boldsymbol\tau}\cdot{\boldsymbol\pi})\right]\psi\;,
\label{lag}
\eqa
where $\psi$ is 
a color $N_c$-plet, a four-component Dirac spinor as well as a flavor doublet 
\bqa
\psi&=&
\left(
\begin{array}{c}
u\\
d
\end{array}\right)\;.
\eqa
Moreover, $\mu_B=3\mu=\mbox{$3\over2$}(\mu_u+\mu_d)$ 
and $\mu_I=(\mu_u-\mu_d)$ 
are the baryon and 
isospin chemical potentials expressed in terms of 
the quark chemical potentials $\mu_u$ and $\mu_d$,
$\tau_i$ ($i=1,2,3$) are the Pauli matrices in flavor space, and 
${\boldsymbol\pi}=(\pi_{1},\pi_{2},\pi_{3})$.

Apart from the global $SU(N_c)$ symmetry, 
the Lagrangian~(\ref{lag}) 
has a 
$U(1)_B\times SU(2)_L\times SU(2)_R$ symmetry for 
$h=0$ and a $U(1)_B\times SU(2)_V$ symmetry
for $h\neq0$. 
When $\mu_u\neq\mu_d$, this symmetry is reduced to 
$U(1)_B\times U_{I_3L}(1)\times U_{I_3R}(1)$ for $h=0$ and
$U(1)_B\times U_{I_3}(1)$ for $h\neq0$. 
In the remainder of this
paper we take $h=0$, i.e. we work in the chiral limit.
We also set $\mu_I=0$.

In the vacuum, the sigma field acquires a nonzero expectation value $\phi_0$.
We can therefore write 
\bqa
\sigma&=&\phi_0+\tilde{\sigma}\;,
\eqa
where $\tilde{\sigma}$ is a quantum fluctuating field with a zero expectation
value. At tree level, the masses of the sigma, the pion, and the quark are
\bqa
\label{m1}
m_{\sigma}^2&=&m^2+{\lambda\over2}\phi_0^2\;,\\
\label{m2}
m_{\pi}^2&=&m^2+{\lambda\over6}\phi_0^2\;,\\
m_q&=&g\phi_0\;.
\label{m3}
\eqa
The tree-level potential $V_{\rm tree}$ is
\bqa
V_{\rm tree}&=&{1\over2}m^2\phi_0^2+{\lambda\over24}\phi_0^4\;,
\eqa
and whose minimum is being identified with the pion decay constant $f_{\pi}$.
The relations (\ref{m1})--(\ref{m3}) can be 
solved with respect to the parameters of the Lagrangian (\ref{lag}).
This yields
\bqa
\label{tr1}
m^2&=&-{1\over2}\left(m_{\sigma}^2-3m_{\pi}^2\right)\;,\\
\lambda&=&3{(m_{\sigma}^2-m_{\pi}^2)\over f_{\pi}^2}\;,\\
g^2&=&{m_q^2\over f_{\pi}^2}\;.
\label{tr4}
\eqa
The Eqs. (\ref{tr1})--(\ref{tr4}) are the parameters determined
at tree level and are often used in practical calculations. However, as
pointed out in the introduction, this is inconsistent in calculations
that involve loop corrections unless one uses the on-shell
renormalization scheme.
In the on-shell scheme, the divergent loop integrals are regularized
using dimensional regularization, but the counterterms are chosen 
differently from the minimal subtraction scheme. 
The counterterms in the on-shell scheme are chosen so that they
exactly cancel the loop corrections to the self-energies and couplings
evaluated on shell, and as a result the
renormalized parameters are independent of the renormalization scale
and satisfy the tree-level relations (\ref{tr1})--(\ref{tr4}).

We need to introduce the counterterms for the parameters in the 
Lagrangian (\ref{lag}), $\delta m^2$, $\delta\lambda$, and $\delta g^2$, the 
wave function counterterms $\delta Z_{\sigma}$,
$\delta Z_{\pi}$, and $\delta Z_{\psi}$. We then write
\bqa
\label{wave1}
\sigma_{\bare}&=&\sqrt{Z_{\sigma}}\sigma\;,
\qquad
\pi_{i\bare}=\sqrt{Z_{\pi}}\pi_i\;,
\\
\psi_{\bare} &=& \sqrt{Z_{\psi}}\psi\;, 
\qquad
m^2_{\bare} = Z_m m^2\;,\\
\lambda_{\bare} &=& Z_\lambda \lambda\;,
\qquad\qquad
g_{\bare}^2 = Z_{g^2} g^2 \;, 
\label{wave4}
\eqa
where $Z_{\sigma}=1+\delta Z_{\sigma}$ etc.
The counterterms $\delta m^2$, $\delta\lambda$, and $\delta g^2$
are expressed in terms of the counterterms
$\delta m_{\sigma}^2$, $\delta m_{\pi}^2$, $\delta m_q$, and $\delta f_\pi^2$.
From Eqs. (\ref{m1})--(\ref{m3}), using Eqs. (\ref{wave1})--(\ref{wave4}),
one finds
\bqa
\delta m^2&=&-{1\over2}\left(\delta m_{\sigma}^2-3\delta m_{\pi}^2\right)
\;,
\label{rela1}
\\
\delta\lambda&=&
3{\delta m_{\sigma}^2-\delta m_{\pi}^2\over f_{\pi}^2}
-\lambda {\delta f_{\pi}^2\over f_{\pi}^2}
\;,
\label{rela2}
\\
\label{rela3}
\delta g^2&=&
{\delta m_q^2\over f_{\pi}^2}
-g^2{\delta f_{\pi}^2\over f_{\pi}^2}
\;.
\eqa
In the large-$N_c$ limit $\delta m_q=0$ and (\ref{rela3}) directly relates 
$\delta g^2$ and $\delta f_\pi^2$. In this limit there are also no loop 
corrections to the pion-quark vertex, which means that the associated 
counterterms must cancel as 
well, leading to $\delta g^2 = -g^2\delta Z_\pi$. Together with (\ref{rela3}) we can rewrite (\ref{rela2}) as

\bqa
\delta\lambda&=&
3{\delta m_{\sigma}^2-\delta m_{\pi}^2\over f_{\pi}^2}
 -\lambda\delta Z_{\pi}
\;.
\eqa

\begin{figure}[htb]
\begin{center}
\includegraphics[width=0.45\textwidth]{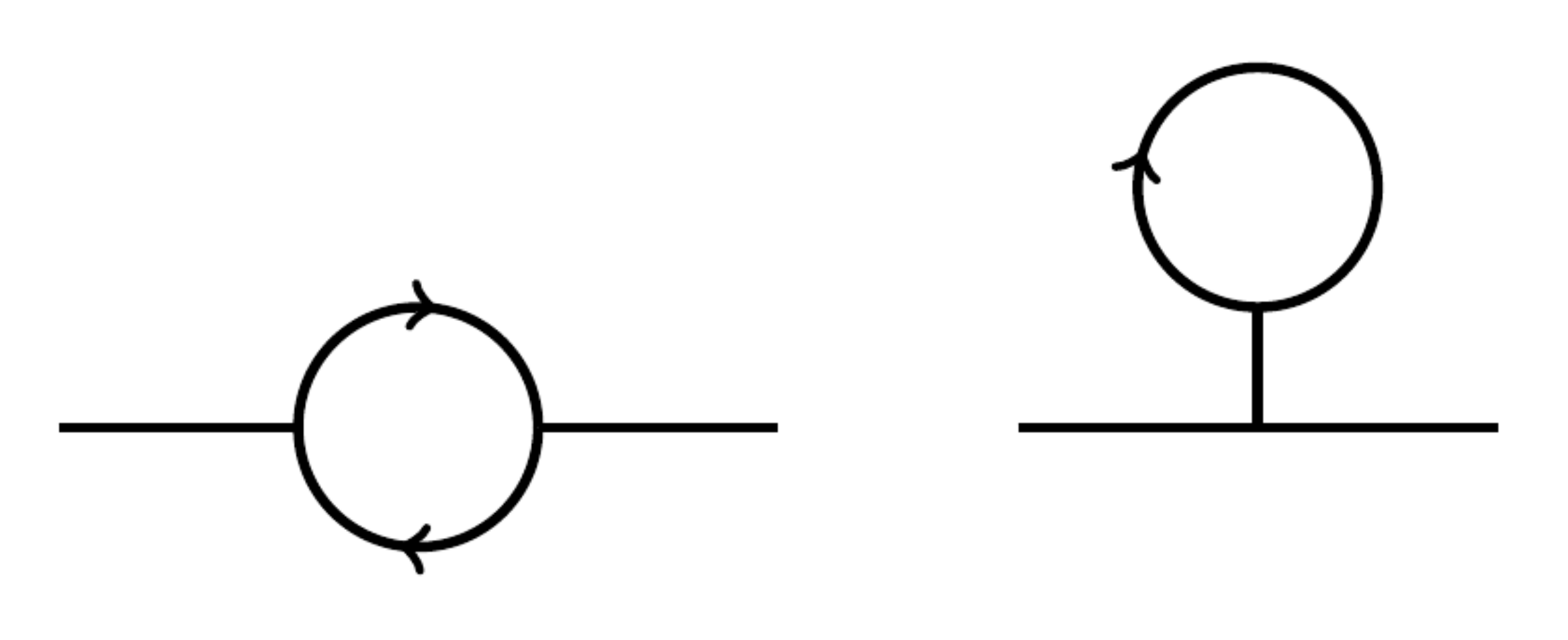}
\end{center}
\caption{One-loop self-energy diagrams for the sigma particle.}
\label{sigmen}
\end{figure}

In the Feynman diagrams below, a 
solid line represents a sigma, a dashed line represents a pion, and the
solid line with an arrow represents a quark.
We work in the large-$N_c$ limit, which implies that we are taking
into account only fermion loops in the self-energies.
The one-loop Feynman diagrams contributing to the 
self-energy of the sigma are shown in Fig. \ref{sigmen}.

The corresponding contributions to the sigma self-energy are given by
\bqa\nonumber
\Sigma_{\sigma}(p^2)&=&
-8g^2N_c
\left[A(m_q^2)-\mbox{$1\over2$}(p^2-4m_q^2)
B(p^2)
\right]
\\ &&
+{4\lambda g \phi_0N_cm_q\over m_{\sigma}^2}A(m_q^2)
\;,
\label{sigsum}
\eqa
where the integrals $A(m^2)$ and $B(p^2)$ are defined in 
Appendix A. 

\begin{figure}[htb]
\begin{center}
\includegraphics[width=0.45\textwidth]{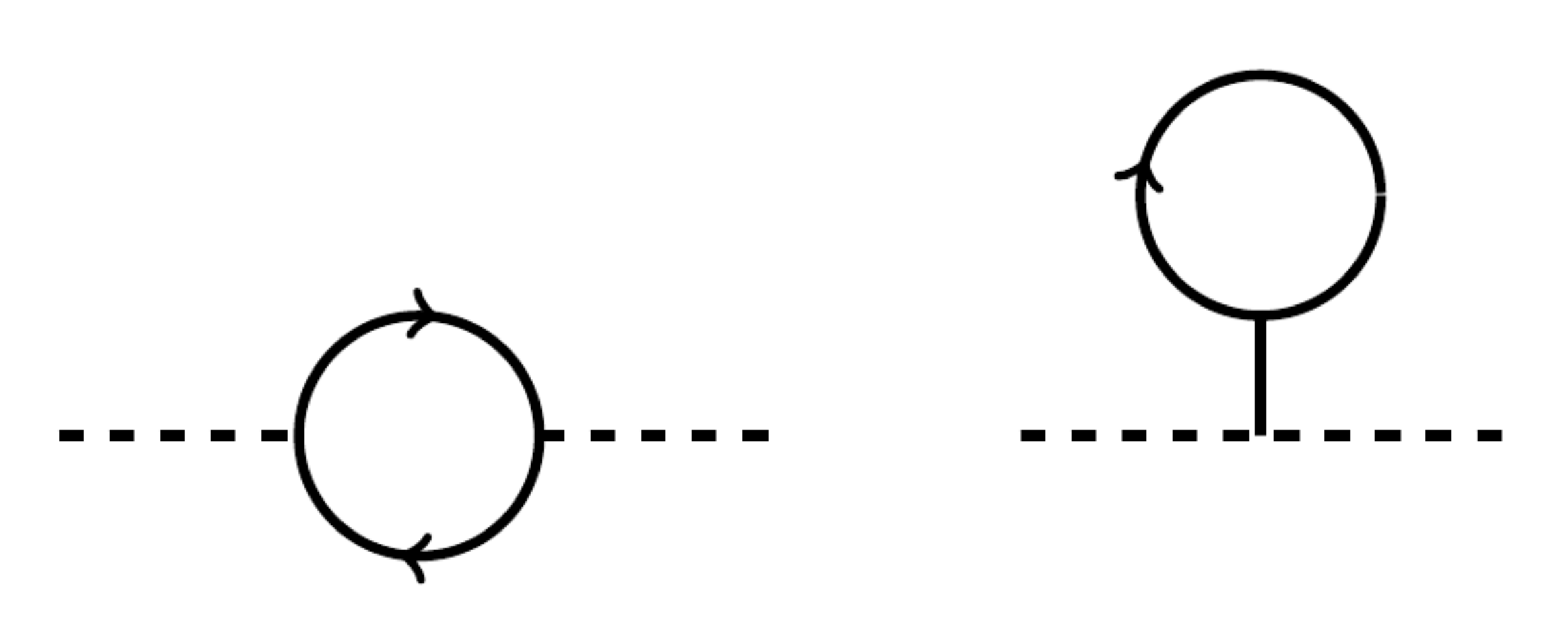}
\end{center}
\caption{One-loop self-energy diagrams for the pion.}
\label{pion}
\end{figure}

The diagrams contributing to the 
self-energy of the pion are shown in Fig. \ref{pion}.

The corresponding contributions to the pion self-energy are given by
\bqa\nonumber
\Sigma_{\pi}(p^2)&=&
-8g^2N_c\left[A(m_q^2)-\mbox{$1\over2$}p^2B(p^2)\right]
\\ 
&&
+{4\lambda g \phi_0N_cm_q\over3m_{\sigma}^2}A(m_q^2)\;.
\label{pisum}
\eqa

\begin{figure}[htb]
\begin{center}
\includegraphics[width=0.45\textwidth]{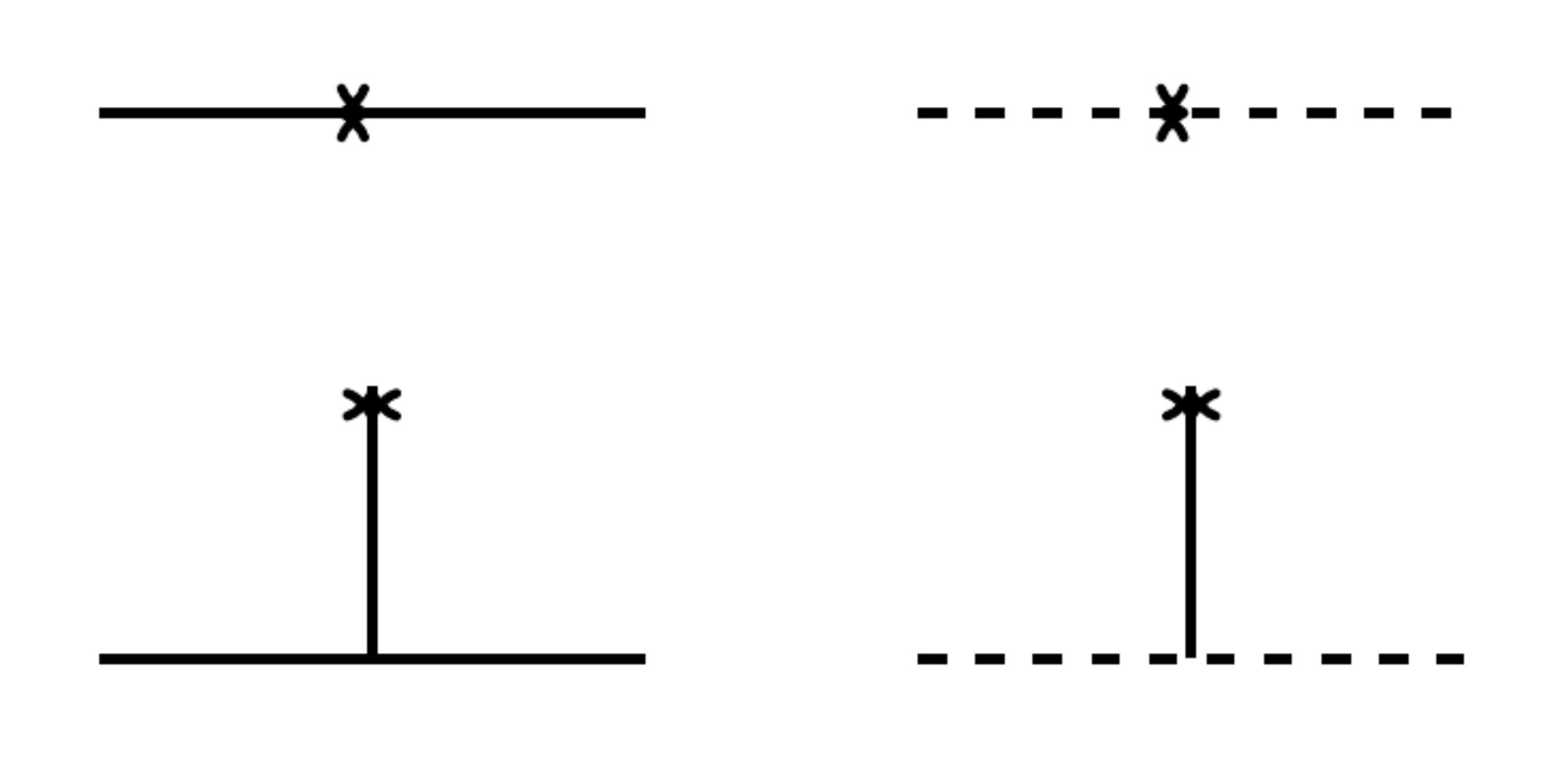}
\end{center}
\caption{Counterterm for the two-point functions for the sigma and pion.}
\label{count11}
\end{figure}

The counterterm diagrams are shown in Fig. \ref{count11}.

We do not need the quark self-energy since it is of order $N_c^0$.
Thus $Z_{\psi}=1$ and $\delta m_q=0$ at this order.

\begin{figure}[htb]
\begin{center}
\includegraphics[width=0.45\textwidth]{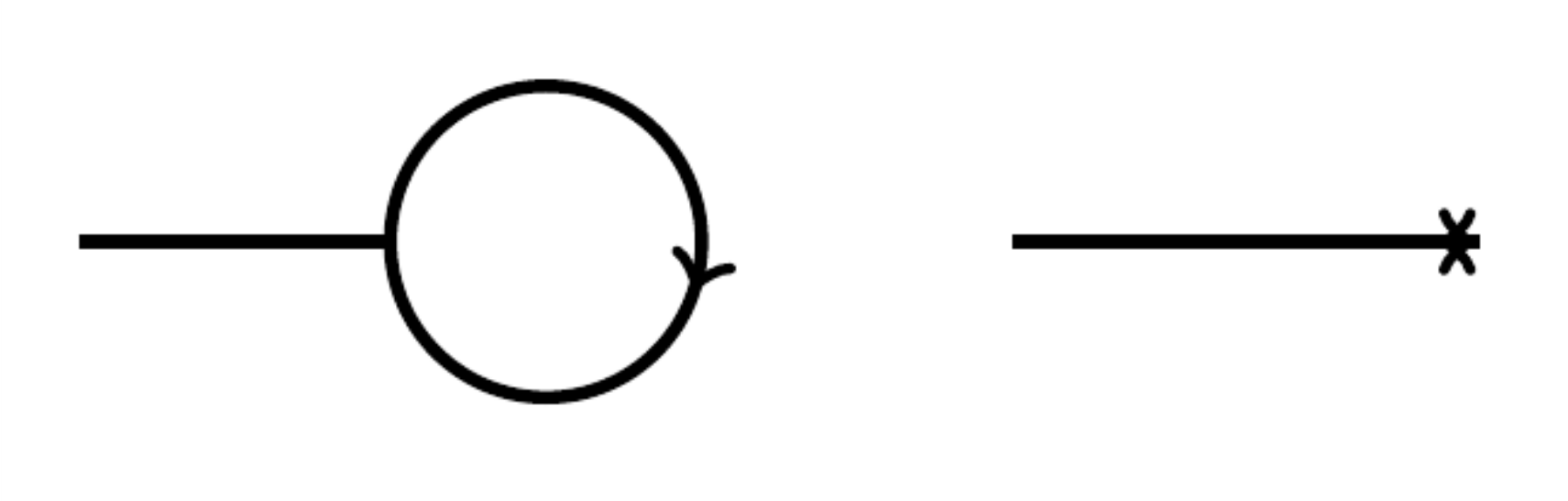}
\end{center}
\caption{Tadpole diagram for the sigma particle and the counterterm.}
\label{onepoint11}
\end{figure}

The one-loop diagram that contributes to the one-point function
together with the counterterm are
shown in Fig. \ref{onepoint11}.

It reads
\bqa
\delta\Gamma^{(1)}
&=&-
8 N_c g m_q A(m_q^2)+i\delta t
\;,
\eqa
where $\delta t$ 
is the counterterm for the tadpole, which can be expressed in terms
of the other counterterms.

\subsection{On-shell renormalization conditions}

The inverse propagator for the sigma or pion 
can be written as
\bqa
p^2-m_{\sigma,\pi}^2-i\Sigma_{\sigma,\pi}(p^2)
{\rm +counterterms}
\;.
\label{definv}
\eqa
In the on-shell scheme, the physical mass 
is equal to the renormalized mass in the Lagrangian, i.e. 
$m=m_{\rm pole}$.\footnote{In defining the mass, we ignore the imaginary parts
of the self-energy.}
Thus we can write
\bqa
\Sigma^{\rm os}(p^2=m_{\sigma,\pi}^2)
{\rm +counterterms}
&=&0
\label{pole}
\;.
\eqa
The residue of the propagator on shell equals unity, which implies
\bqa
\label{res}
{\partial\over\partial p^2}\Sigma_{\sigma,\pi}(p^2)\Big|_{p^2=m_{\sigma,\pi}^2}
{\rm +counterterms}
&=&0\;.
\eqa
The equation of motion is that the one-point function vanishes.
At tree level, the equation of motion is
$t=m_{\pi}^2\phi_0=0$, and in the broken phase the pion mass 
is zero in accordance with Goldstone's theorem. 
The renormalization condition is then
\bqa
\delta\Gamma^{(1)}
&=&0\;.
\eqa
The counterterms that correspond to Figs. \ref{count11}
and \ref{onepoint11} are given by
\bqa
\label{count1}
\Sigma_{\sigma}^{\rm ct}(p^2)
&=&i\left[
\delta Z_{\sigma}(p^2-m_{\sigma}^2)-\delta m_{\sigma}^2
\right]\;,\\
\label{count2}
\Sigma_{\pi}^{\rm ct}(p^2)
&=&i\left[
\delta Z_{\pi}(p^2-m_{\pi}^2)-\delta m_{\pi}^2
\right]\;,\\
\Sigma_{\sigma}^{\rm ct2}&=&3\Sigma_{\pi}^{\rm ct2}
={i\lambda\phi_0^2\over2m_{\sigma}^2}\delta m_{\pi}^2\;,
\\
\delta t&=&
-
\phi_0\delta m_{\pi}^2
\;.
\eqa
The on-shell renormalization constants are given by the
self-energies and their derivatives 
evaluated on shell. Combining Eqs. (\ref{pole})--(\ref{count2}), we find
\bqa
\delta m_{\sigma}^2&=&-i\Sigma_{\sigma}(m_{\sigma}^2)\;,\\
\delta m_{\pi}^2&=&-i\Sigma_{\pi}(0)\;,\\
\delta Z_\sigma &=&
i{\partial\over\partial p^2}\Sigma_\sigma(p^2)\Big|_{p^2=m_\sigma^2}\;, \\
\delta Z_\pi&=& i{\partial\over\partial p^2}\Sigma_\pi(p^2)\Big|_{p^2=m_\pi^2}\;.
\eqa
From Eqs. (\ref{sigsum}) and (\ref{pisum}), we find
\footnote{The self-energies are without the tadpole contributions.}
\bqa
\delta m_{\sigma}^2
&=&
8ig^2N_c\left[A(m_q^2)-\mbox{$1\over2$}(m_{\sigma}^2-4m_q^2)B(m_{\sigma}^2)
\right]\;,
\\ 
\delta m_{\pi}^2&=&8ig^2N_cA(m_q^2)\;,
\\
\delta Z_{\sigma}&=&
4ig^2N_c\left[B(m_{\sigma}^2)+(m_{\sigma}^2-4m_q^2)B^{\prime}(m_{\sigma}^2)
\right]\;,
\\
\delta Z_{\pi}&=&4ig^2N_cB(0)\;.
\eqa
Using Eqs. (\ref{rela1})--(\ref{rela3}), we find expressions
for the counterterms $\delta m_{\os}^2$, $\delta\lambda_{\os}$, 
and $\delta g^2_{\os}$,
\begin{figure}[htb]
\end{figure}
\begin{widetext}
\bqa\nonumber
\delta m_{\os}^2&=&
8ig^2N_c\left[A(m_q^2)+\mbox{$1\over4$}(m_{\sigma}^2-4m_q^2)B_0(m_{\sigma}^2)
\right]
\\&=&=
\delta m^2_{\rm div} + m^2\dfrac{4g^2 N_c}{(4\pi)^2} 
\left[ \log\left(\dfrac{\Lambda^2}{m_q^2}\right) +\dfrac{4m_q^2}{m_\sigma^2} 
+\left( 1-\dfrac{4m_q^2}{m_\sigma^2}\right) F(m_\sigma^2) \right]  
\;,
\label{osse1}
\\ \nonumber
\delta\lambda_{\os}&=&
-\dfrac{12ig^2N_c}{f_{\pi}^2} (m_\sigma^2-4m_q^2)B(m_\sigma^2)
-4i\lambda g^2N_c B(0) 
\\ 
&=&
\delta\lambda_{\rm div} + \dfrac{12g^2N_c}{(4\pi)^2}{m_\sigma^2\over f_\pi^2}
\left[ \left( 2 -\dfrac{4m_q^2}{m_\sigma^2} \right)\log\left(\dfrac{\Lambda^2}{m_q^2}\right) +\left( 1 -\dfrac{4m_q^2}{m_\sigma^2} \right)F(m_\sigma^2)
\right]
\;,
\label{osse2}
\\ 
\delta g^2_{\os}&=&
-4ig^4N_c B(0)
=\delta g_{\rm div}^2
+\dfrac{4g^4 N_c}{(4\pi)^2} \log\left(\dfrac{\Lambda^2}{m_q^2}\right)  
\;,
\label{osse3}
\\
\delta Z_{\sigma}^{\os}
&=&
\delta Z_{\sigma,\rm div}
-\dfrac{4g^2 N_c}{(4\pi)^2}\left[ \log\left(\dfrac{\Lambda^2}{m_q^2}\right)  
+F(m_\sigma^2) +(m_\sigma^2-4m_q^2)F^\prime(m_\sigma^2) \right]
\;\;\;, \\
\delta Z_{\pi}^{\os}
&=&
\delta Z_{\pi,\rm div}
-\dfrac{4g^2 N_c}{(4\pi)^2} \log\left(\dfrac{\Lambda^2}{m_q^2}\right)
\;\;\;,
\eqa
\end{widetext}
where $F(m^2)$ and $F^{\prime}(m^2)$
are defined in Appendix A, 
and the divergent quantities are
\bqa
\delta m^2_{\rm div}&=&
m^2{4g^2N_c \over(4\pi)^2\epsilon}
\;,\\
\delta\lambda_{\rm div}&=&
{8g^2N_c\over(4\pi)^2\epsilon}\left(\lambda
-6g^2\right)
\;,\\
\delta g_{\rm div}^2&=&{4g^4N_c\over(4\pi)^2\epsilon}\;,\\
\delta Z_{\sigma,\rm div}&=&\delta Z_{\pi,\rm div}=-{4g^2N_c\over(4\pi)^2\epsilon}
\;.
\eqa
The divergent parts of the counterterms 
are the same in the two schemes, i.e.
$\delta m^2_{\rm div}=\delta m^2_{\ms}$ and so forth.
\section{Relations between parameters in the two schemes}
Since the bare parameters are independent
of the renormalization scheme, we can immediately
write down the relations between
the renormalized parameters in the on-shell and $\overline{\rm MS}$ schemes.
We find
\bqa
m^2_{\ms} &=& 
\dfrac{Z_m^{\os}}{Z_m^{\ms}}m^2 \approx m^2 + \delta m^2_{\os} 
- \delta m^2_{\ms} \;,
\\
\lambda_{\ms} &=& 
\dfrac{Z_\lambda^{\os}}{Z_\lambda^{\ms}}\lambda \approx \lambda + \delta\lambda_{\os} 
 - \delta\lambda_{\ms} \;,
\\
g_{\ms}^2 &=& \dfrac{Z_{g^2}^{<\os}}{Z_{g^2}^{\ms}}g^2 \approx g^2 + \delta g^2_{\os}
- \delta g^2_{\ms}\;.
\eqa
Using Eqs. (\ref{osse1})--(\ref{osse3}), we find the 
running parameters in the $\overline{\rm MS}$ scheme
\begin{widetext}
\bqa \nonumber
m^2_{\ms} 
&=& m^2 +8ig^2N_c \left[ A(m_q^2) +\mbox{$1\over4$}(m_\sigma^2-4m_q^2)B(m_\sigma^2)
\right] -\delta m^2_{\ms} \\ 
 &=& -\mbox{$1\over2$}m_\sigma^2\left\{ 1 +\dfrac{4m_q^2N_c}{(4\pi)^2 f_\pi^2} 
\left[ \log\left(\dfrac{\Lambda^2}{m_q^2}\right) 
+\dfrac{4m_q^2}{m_\sigma^2} 
+\left(1-\dfrac{4m_q^2}{m_\sigma^2}
\right)F(m_\sigma^2) 
\right] \right\}
\;, 
\label{osm1}
\\ \nonumber
\lambda_{\ms} 
&=& \lambda -4i\lambda g^2N_c
\left[ \left( 1-\dfrac{4m_q^2}{m_\sigma^2}  \right) B(m_\sigma^2) 
+B(0)
\right] 
- \delta\lambda_{\ms}
\\
 &=& \dfrac{3 m_\sigma^2}{f_\pi^2} \left\{1 +\dfrac{4m_q^2N_c}{(4\pi)^2 f_\pi^2}
\left[ \left( 2 -\dfrac{4m_q^2}{m_\sigma^2} \right)\log\left(\dfrac{\Lambda^2}{m_q^2}\right) +\left( 1 -\dfrac{4m_q^2}{m_\sigma^2} \right)F(m_\sigma^2)
\right] \right\}\;,
\label{osl}
\\ \nonumber
g_{\ms}^2 
 &= &g^2-4ig^4N_c B(0) -\delta g_{\ms}^2 
\\ 
 &=& {m_q^2\over f_{\pi}^2}
\left\{1 + \dfrac{4m_q^2N_c}{(4\pi)^2 f_\pi^2} \log\left(\dfrac{\Lambda^2}{m_q^2}\right) \right\}
\;,
\label{osm3}
\eqa
\end{widetext}
where the physical on-shell values are related to the
meson and quark masses given by Eqs. (\ref{tr1})--(\ref{tr4}).

The running parameters $m_{\ms}^2(\Lambda)$, 
$\lambda_{\ms}(\Lambda)$, and $g_{\ms}^2(\Lambda)$
satisfy a set of renormalization group equations, which in the large-$N_c$ limit are
\bqa
\label{run1}
\Lambda{dm_{\ms}^2(\Lambda)\over d\Lambda}&=&
{8m_{\ms}^2(\Lambda)g^2_{\ms}(\Lambda)N_c\over(4\pi)^2}\;,
\\
\Lambda{dg^2_{\ms}(\Lambda)\over d\Lambda}
&=&{8g_{\ms}^4(\Lambda)N_c\over(4\pi)^2}\;,
\\
\Lambda{d\lambda_{\ms}(\Lambda)\over d\Lambda}&=&{16N_c\over(4\pi)^2}
\left[
\lambda_{\ms}(\Lambda)g_{\ms}^2(\Lambda)-6g_{\ms}^4(\Lambda)\right]
\;,
\label{run3}
\eqa
The solutions to Eqs. (\ref{run1})--(\ref{run3}) are
\bqa
\label{sol1}
m_{\ms}^2(\Lambda)&=&{m_0^2\over1-{4g_0^2N_c\over(4\pi)^2}
\log{\Lambda^2\over m_q^2}}\;,
\\
g_{\ms}^2(\Lambda)&=&
{g_0^2\over1-{4g_0^2N_c\over(4\pi)^2}
\log{\Lambda^2\over m_q^2}
}\;,
\\
\label{sol4}
\lambda_{\ms}(\Lambda)&=&{\lambda_0-{48g_0^4N_c\over(4\pi)^2}
\log{\Lambda^2\over m_q^2}
\over\left(1-{4g_0^2N_c\over(4\pi)^2}
\log{\Lambda^2\over m_q^2}
\right)^2}\;,
\eqa
where $m_0^2$, $g_0^2$, and $\lambda_0$ are the 
values of the running mass and couplings at the scale 
$\Lambda=m_q$. They are found by evaluating 
Eqs. (\ref{osm1})--(\ref{osm3}) at this scale.

In the Nambu-Jona-Lasinio model, we have the relation $m_{\sigma}=2m_q$ \cite{klev}, while
there is no such relation between the sigma mass and quark mass
in the quark-meson model. However, it is interesting to note that 
for $m_\sigma=2m_q$, the tree-level relation $\lambda=12g^2$
is valid at the one-loop level in the large-$N_c$ limit;
using $\lambda_0=3{m_{\sigma}^2\over f_\pi^2}=12g_0^2$, we find 
$\lambda_{\ms}(\Lambda)=12g_{\ms}^2(\Lambda)$.

\section{Results and Discussion}
In this section, we calculate the one-loop effective potential
and study the phase diagram.
We are working in the large-$N_c$ limit, 
which implies that only fermion loops are taken into account. 
This is often referred to as
the mean-field approximation. The one-loop contribution to the
effective potential
is straightforward to calculate in this limit and reads
\bqa
V_{1}&=&-4N_c\sumint_{\{P\}}\log\left[P^2+\Delta^2\right]\;,
\eqa
where the sum-integral 
is defined in Appendix A.
After redefining the field $\phi_0$ and renormalizing the 
mass parameter $m^2$ and coupling constants $g^2$
and $\lambda$, we find 
\begin{widetext}
\bqa\nonumber
V_{\rm 1-loop}&=&{1\over2}m_{\ms}^2(\Lambda){\Delta^2\over g_{\ms}^2(\Lambda)}
+{\lambda_{\ms}(\Lambda)\over24}{\Delta^4\over g_{\ms}^4(\Lambda)}
+{2N_c\Delta^4\over(4\pi)^2}\left[
\log{\Lambda^2\over\Delta^2}+{3\over2}
\right]
\\ &&
-4N_cT\int_p
\bigg\{
\log\left[1+e^{-\beta(E-\mu)}\right]
+\log\left[1+e^{-\beta(E+\mu)}\right]\bigg\}\;,
\label{pot111}
\eqa
where 
$\mu=\mu_u=\mu_d$ is the quark chemical potential, 
and $E=\sqrt{p^2+\Delta^2}$.
Substituting the running parameters Eqs. 
(\ref{sol1})--(\ref{sol4}) 
into Eq. (\ref{pot111}), the effective potential becomes
independent of the renormalization scale $\Lambda$ and reads
\bqa\nonumber
V_{\rm1-loop}&=& -\dfrac{1}{4}m_\sigma^2 f_\pi^2
\left\{
1 +\dfrac{4 m_q^2N_c}{(4\pi)^2f_\pi^2}\left[ \left( 1 -\dfrac{4m_q^2}{m_\sigma^2}
\right)F(m_\sigma^2)
 +\dfrac{4m_q^2}{m_\sigma^2}
\right]\right\}\dfrac{\Delta^2}{m_q^2} \\ \nonumber
 & & + \dfrac{1}{8}m_\sigma^2 f_\pi^2
\left\{ 1 -\dfrac{4 m_q^2  N_c}{(4\pi)^2f_\pi^2}\left[
\dfrac{4m_q^2}{m_\sigma^2}\left( \log\left( \dfrac{\Delta^2}{m_q^2} \right) 
-\dfrac{3}{2} \right) 
-\left( 1 -\dfrac{4m_q^2}{m_\sigma^2}\right)F(m_\sigma^2)
\right]
 \right\}\dfrac{\Delta^4}{m_q^4} \\
&&-4N_cT\int_p
\bigg\{
\log\left[1+e^{-\beta(E-\mu)}\right]
+\log\left[1+e^{-\beta(E+\mu)}\right]\bigg\}\;.
\eqa
\end{widetext}
In the remainder of the paper, we set $N_c=3$.
Moreover, the mass of the sigma particle is not known 
very accurately \cite{databook}.
It is therefore common to vary it within the range of 400--800 MeV to
study the effects on the phase diagram.

\begin{figure}[htb]
\begin{center}
\includegraphics[width=0.45\textwidth]{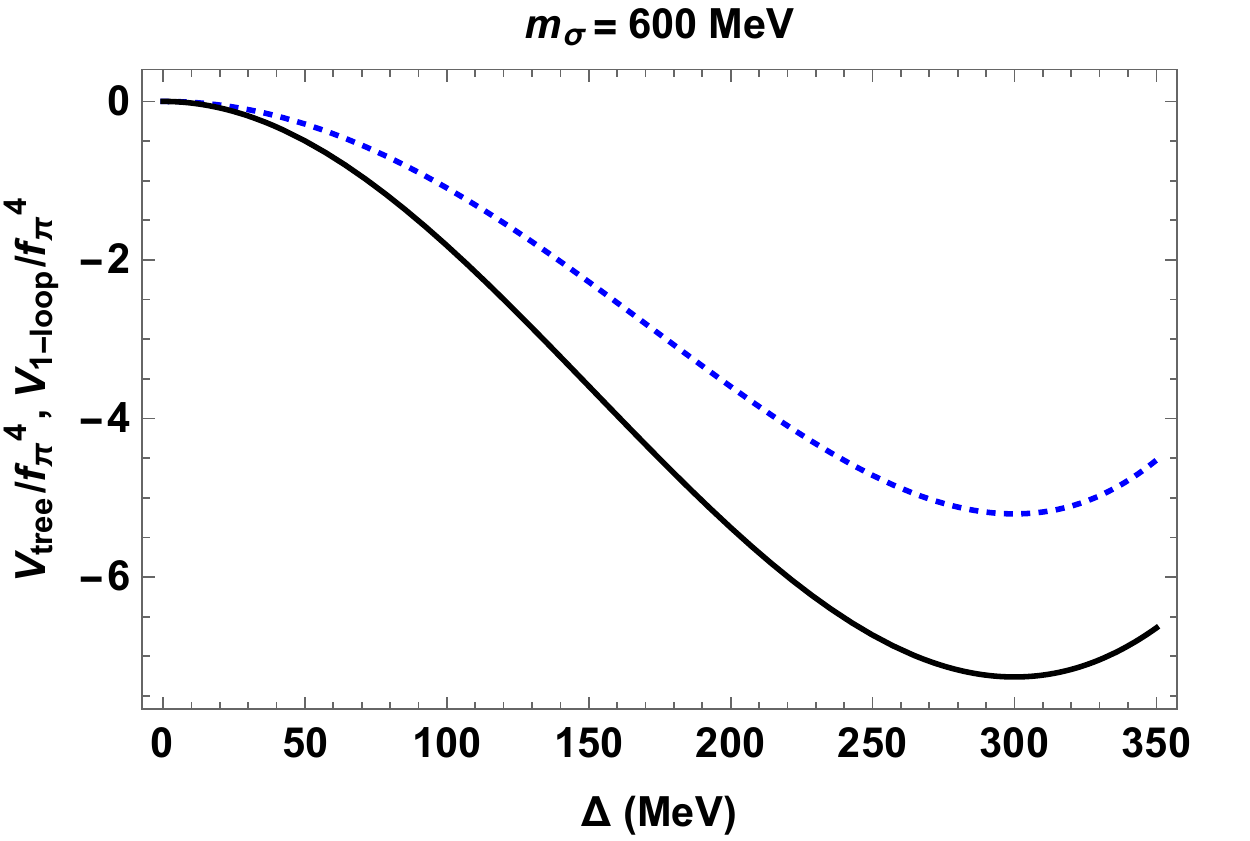}
\end{center}
\caption{Vacuum effective potential normalized to $f_{\pi}^4$
as a function of $\Delta$ for $m_{\sigma}=600$ MeV. 
Dashed line is the tree-level potential
and the solid line is the one-loop effective potential in the 
large-$N_c$ limit.}
\label{potential}
\end{figure}

In Fig. \ref{potential}, we show the normalized
tree-level (dashed line)
as well as the one-loop (solid line)
effective potential in the vacuum ($\mu=T=0$)
as a function of $\Delta$ for $m_{\sigma}=600$ MeV.
This corresponds to the NJL relation between the sigma mass
and the constituent quark mass, $m_{\sigma}=2m_q$.
Both potentials have a minimum at $\Delta=300$ MeV, 
but the one-loop effective potential is significantly deeper.

\begin{figure}[htb]
\begin{center}
\includegraphics[width=0.45\textwidth]{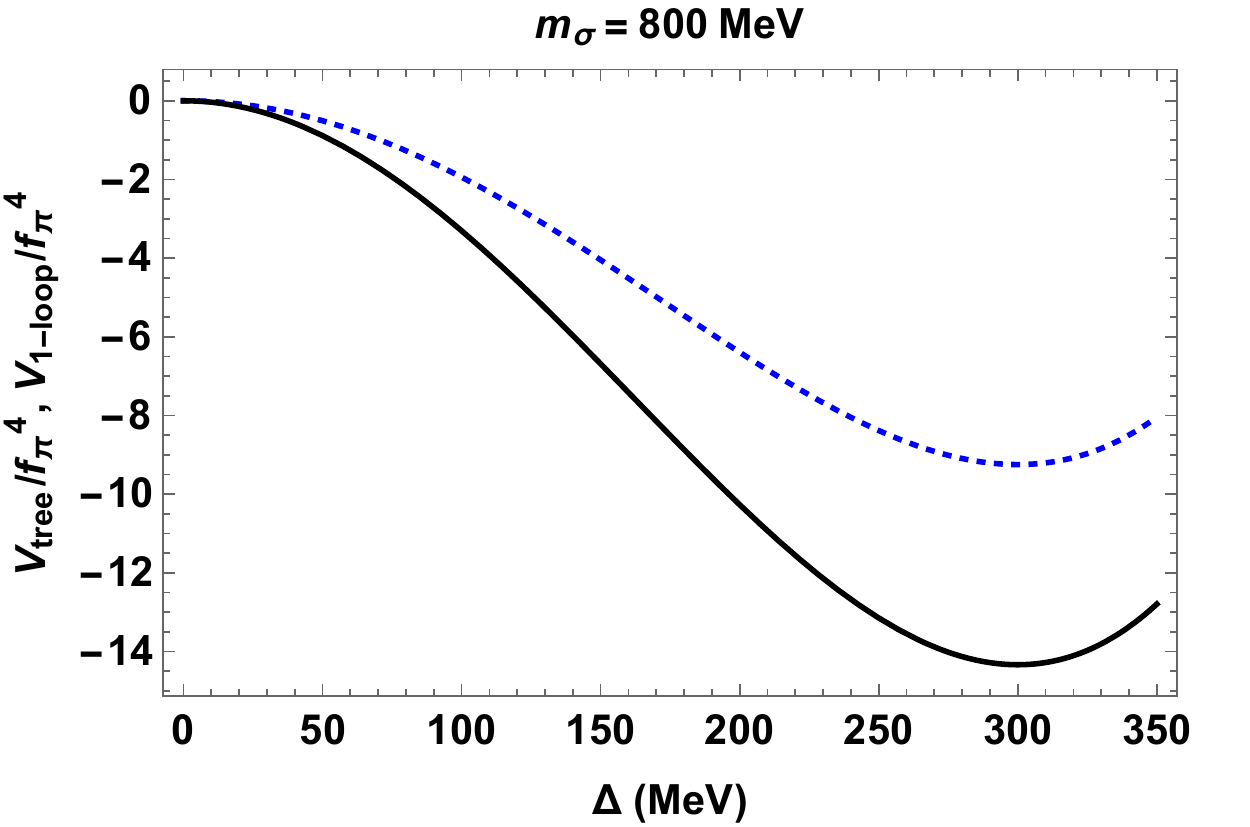}
\end{center}
\caption{Vacuum effective potential normalized to $f_{\pi}^4$
as a function of $\Delta$ for $m_{\sigma}=800$ MeV.
Dashed line is the tree-level potential
and the solid line is the one-loop effective potential in the 
large-$N_c$ limit.}
\label{potential2}
\end{figure}

In Fig. \ref{potential2}
we also show the normalized
tree-level (dashed line)
as well as the one-loop (solid line)
effective potential in the vacuum ($\mu=T=0$)
as a function of $\Delta$ for $m_{\sigma}=800$ MeV. 
Qualitatively, the potential looks the same as in Fig. \ref{potential}.

\begin{figure}[htb]
\begin{center}
\includegraphics[width=0.45\textwidth]{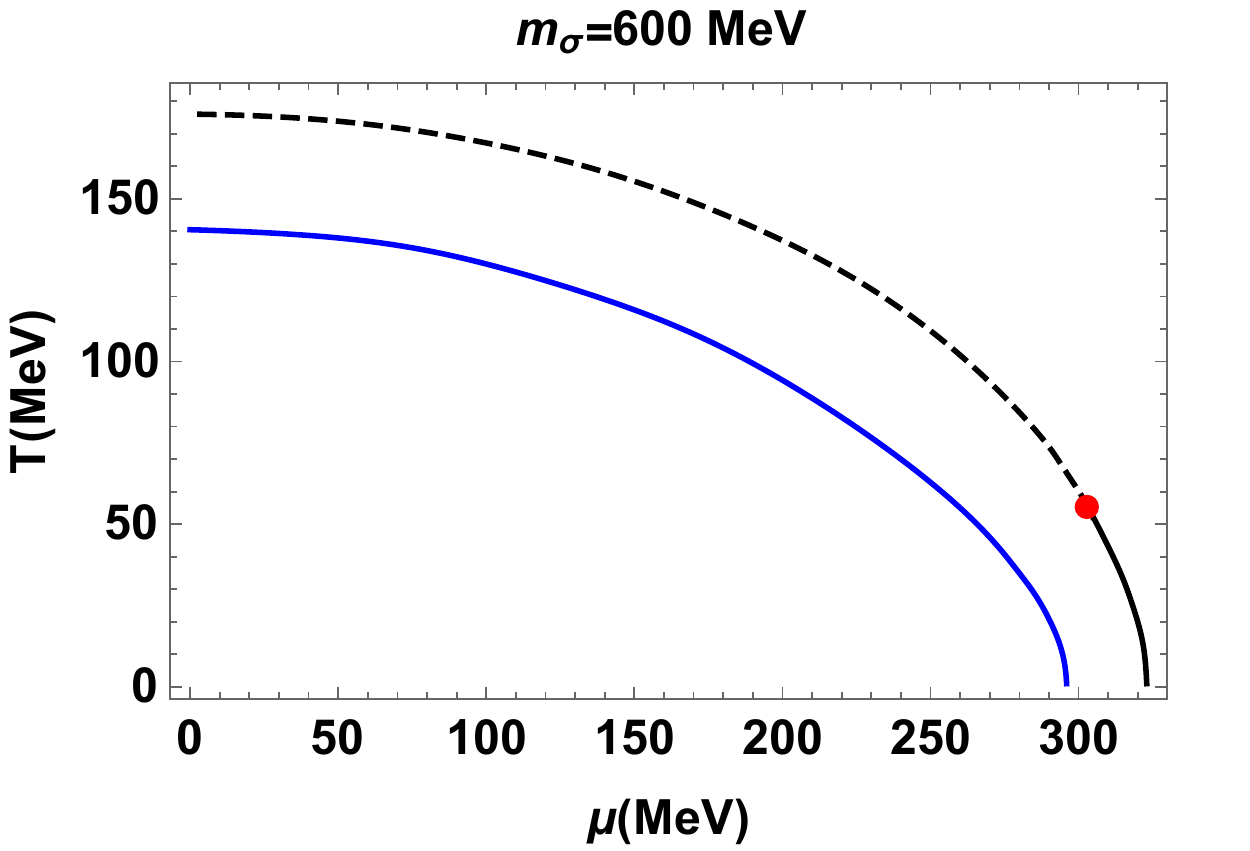}
\end{center}
\caption{The phase diagram in the $\mu$--$T$ plane for $m_{\sigma}=600$ MeV.
A dashed line indicates a second-order transition, while a 
solid indicates a first-order transition. The red dot shows the
tricritical point.
The blue solid line is
phase boundary in the no-sea approximation.}
\label{600}
\end{figure}

In Fig. \ref{600}, we show the 
phase diagram in the $\mu$--$T$ plane for $m_{\sigma}=600$ MeV.
If one excludes the vacuum fluctuations of the fermions and hence
ignores renormalization issued altogether, the model predicts
a first-order transition in the entire $\mu$--$T$ plane.
For vanishing chemical baryon potential $\mu$, 
universality arguments suggest that it is second order \cite{rob},  
and strongly suggests that
one should take the vacuum fluctuations of any model 
seriously \cite{vac,lars,gupti}.
Moreover, the first-order transition that starts at $T=0$, ends
at the tricritical point
indicated by a red dot and located at
$(\mu,T) = (303.24 \; \text{MeV}, 55 \; \text{MeV})$.

\begin{figure}[htb]
\begin{center}
\includegraphics[width=0.45\textwidth]{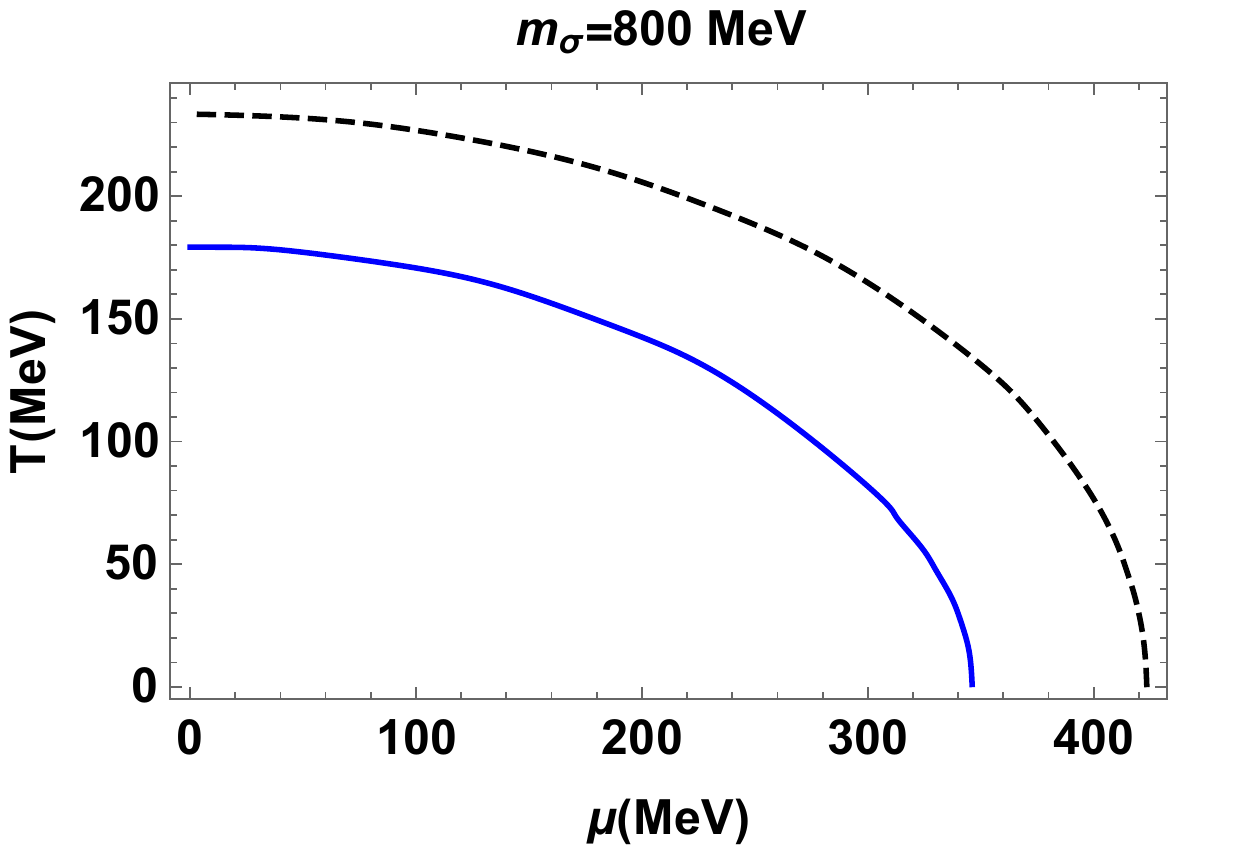}
\end{center}
\caption{The phase diagram in the $\mu$--$T$ plane for $m_{\sigma}=800$ MeV.
A dashed line indicates a second-order transition, while a 
solid indicates a first-order transition. The blue solid line is the
phase boundary in the no-sea approximation.}
\label{800}
\end{figure}

In Fig. \ref{800}, we show the 
phase diagram in the $\mu$--$T$ plane for $m_{\sigma}=800$ MeV.
The transition is now of second order in the entire $\mu$--$T$ plane, 
if one includes vacuum fluctuations and first order
if they are neglected.
For both values of $m_{\sigma}$, the critical temperature increases
significantly by including the vacuum fluctuations
and one-loop corrections to the parameters of the Lagrangian (\ref{lag}).
Our results for a sigma mass of 600 MeV and 800 MeV are 
in very good agreement with those of Ref. \cite{fix1}, 
where the authors use Pauli-Villars regularization and the pole mass
definition to study the phase diagram of the QM model.
There are a number of other studies of this model; however,
a quantitative comparison is difficult since the curvature of the effective
potential is used to define the sigma mass 
(see discussion below)
or because matching is done
at tree level. Qualitatively, the functional-renormalization group 
(FRG) study
in Ref. \cite{frg} predicts a more complicated phase structure at low
$T$. The second-order line starting at $\mu=0$ ends at a tricritical point.
The first-order transition bifurcates at larger values of $\mu$
where one of the branches is first order, while the second branch initially
is first order and then second order. This more complicated structure
may very well be related to the fact that the FRG includes mesonic 
fluctuations.

A common, but incorrect definition of the sigma mass
is the second derivative of the effective potential
in the minimum. This is often referred to as the curvature mass.
The effective potential is the generator of the $n$-point functions
of the theory at vanishing external momenta and so the curvature
mass is equivalent to defining the sigma mass using the 
self-energy evaluated at vanishing momentum.
The difference between the two masses is finite, but a priori 
difficult to quantify. In hot gauge theories, the correct way
of defining the mass has a long history, and we emphasize that
the pole definition is the physical and gauge invariant one
\cite{rebhan1,rebhan2}.
If different definitions of masses 
are used or if tree-level relations are applied at the loop level,
one cannot compare different model predictions quantitatively.
It is therefore important to determine the parameters in the
Lagrangian in the correct way.

To summarize, we have calculated the running parameters 
$m^2$, $\lambda$, and $g^2$ at one loop 
by relating the $\overline{\rm MS}$ and on-shell schemes
and the experimental values for the meson and quark masses 
and pion decay constant.
We used this as input 
to the one-loop effective potential 
that was used to map out the phase diagram in the $\mu$--$T$ plane.
We will present a more complete analysis including the Polyakov loop
variable $\Phi$ and the possibility of inhomogeneous phases in a
forthcoming publication \cite{forth}.
The correct determination of the parameters in the 
quark-meson model should be useful in other contexts.
For example, the $SU(3)$ 
quark-meson model has been used to
study the phase diagram of QCD and quark stars.

\begin{figure}
\end{figure}

\section*{Acknowledgments}
The authors would like to thank the Niels Bohr International Acedemy
for its hospitality during the early stages of this work.
J.O.A. would like to thank S. Carignano and M. Buballa for useful
discussions.

\section*{INTEGRALS AND SUM INTEGRALS}
The divergent loop integrals are regularized using dimensional
regularization. We define the dimensionally regularized integrals by
\bqa
\int_p&=&\left({e^{\gamma_E}\Lambda^2\over4\pi}\right)^{\epsilon}
\int{d^Dp\over(2\pi)^D}\;,
\eqa
where $D=4-2\epsilon$, $\gamma_E$ is the Euler-Mascheroni constant, and
$\Lambda$ is the renormalization scale associated with the 
$\overline{\rm MS}$ scheme.
Specifically, we need the integrals
\bqa\nonumber
A(m^2)&=&
\int_p{1\over p^2-m^2}
\\&=& 
{im^2\over(4\pi)^2}\left({\Lambda^2\over m^2}\right)^{\epsilon}
\left[{1\over\epsilon}+1\right]\;,
\\ \nonumber
B(p^2)&=&
\int_k{1\over(k^2-m^2)[(k+p)^2-m^2]}
\\ &=& 
{i\over(4\pi)^2}\left({\Lambda^2\over m^2}\right)^{\epsilon}\left[
{1\over\epsilon}+F(p^2)
\right]\;,
\\
B^{\prime}(p^2)&=&{i\over(4\pi)^2}F^{\prime}(p^2)\;,
\eqa
where the functions
$q$, $F(p^2)$, and $F^{\prime}(p^2)$ are
\bqa
q&=&\sqrt{{4m^2\over p^2}-1}\;,
\\ \nonumber
F(p^2)&=&
-\int_0^1dx\,\log\left[{p^2\over m^2}x(x-1)+1\right]
\\&=&
2-2q\,{\arctan}\left(\mbox{$1\over q$}\right)
\;,
\\
F^{\prime}(p^2)&=&
{4m^2q\over p^2(4m^2-p^2)}{\arctan}
\left(\mbox{$1\over q$}\right)
-{1\over p^2}
\;.
\eqa

In the imaginary-time formalism for thermal field theory, a 
fermion has Euclidean 4-momentum $P=(P_0,{\bf p})$ with
$P^2=P_0^2+{\bf p}^2$. The Euclidean energy $P_0$
has discrete values:
$P_0=(2n+1)\pi T+i\mu$, where $n$ is an
integer. Loop diagrams involve a sum over
$P_0$ and an integral over spatial momenta 
$p$. With dimensional regularization,
the integral is generalized to $d=3-2\epsilon$
spatial dimensions. We define the dimensionally regularized
sum-integral by
\bqa
\sumint_{\{P\}}&=&
T\sum_{\{P_0\}}
\int_p
\;,
\label{sumint}
\eqa
where $\Lambda$ is the renormalization scale in the
modified minimal subtraction scheme $\overline{\rm MS}$ and
\bqa
\int_p&=&
\left({e^{\gamma_E}\Lambda^2\over4\pi}\right)^{\epsilon}
\int{d^{d}p\over(2\pi)^d}\;.
\label{dint}
\eqa
Specifically, we need the sum-integral
\bqa
I_0&=&\sumint_{\{P\}}\log\left[P^2+m^2\right]\;.
\eqa
Summing over the Matsubara frequencies $P_0$, we obtain
\bqa\nonumber
I_0&=&-\int_p\sqrt{p^2+m^2}
-T\int_p\bigg\{\log\left[1+e^{-\beta(E-\mu)}\right]
\\ &&
+\log\left[1+e^{-\beta(E+\mu)}\right]\bigg\}\;.
\eqa
The first term is ultraviolet divergent and in dimensional regularization
it reads
\bqa
\int_p\sqrt{p^2+m^2}
&=&-{2m^4\over(4\pi)^2}\left(
{\Lambda^2\over m^2}
\right)^{\epsilon}\left[
{1\over\epsilon}+{3\over2}
\right]\;.
\eqa


\bibliography{refs}{}

\begin{thebibliography}{99}
\bibitem{rev1}
M. G. Alford, A. Schmitt, and K .Rajagopal, 
Rev. Mod. Phys. {\bf 80}, 1455 (2008), 
\bibitem{rev2}
K. Fukushima, and T. Hatsuda,
Rep. Prog. Phys. {\bf 74}, 014001 (2011).

\bibitem{fuku}
K. Fukushima, 
Phys. Lett. B {\bf 591}, 277 (2004).

\bibitem{benji}
B. Svetitsky and L. G. Yaffe,
Nucl. Phys. B {\bf 210}, 423 (1982).



\bibitem{ratti}C. Ratti, M. A. Thaler, and W. Weise, 
Phys. Rev. D {\bf 73},
014019 (2006).

\bibitem{fuku2}
K. Fukushima, Phys. Rev. D {\bf 78}, 114019 (2008).

\bibitem{scav}	
O. Scavenuius, A. Mocsy, I. N. Mishustin, and D. H. Rischke,
Phys. Rev. C {\bf 64}, 045202 (2001).





 	
\bibitem{sirlin1} A. Sirlin, Phys. Rev. D {\bf 22}, 971 (1980).
\bibitem{sirlin2} A. Sirlin, Phys. Rev. D {\bf 29}, 89 (1984).

\bibitem{hollik}
M. Bohm, H. Spiesberger, and W. Hollik,
Fortsch. Phys. {\bf 34},  687 (1986).
\bibitem{hollik2}
W Hollik, Fortsch. Phys. {\bf 38}, 165, (1990).


\bibitem{laine}
K. Kajantie, M. Laine, K. Rummukainen, and M. E. Shaposhnikov
Nucl. Phys. B {\bf 458}, 90 (1996).

\bibitem{chiku0}	
S. Chiku and T. Hatsuda,
Phys. Rev. D {\bf 57}, 6 (1998).

\bibitem{chiku}
S. Chiku and T. Hatsuda,
Phys. Rev. D {\bf 58}, 076001 (1998).

\bibitem{hidaka}
Y. Hidaka, O. Morimatsu, and T. Nishikawa,
Phys. Rev. D {\bf 67}, 056004 (2003).

\bibitem{fix1}
S. Carignano, M. Buballa, and W. Elkamhawy,
Phys. Rev. D {\bf 94}, 034023 (2016).


\bibitem{databook}
Particle data group,
http://pdg.lbl.gov/2014/listings/rpp2014-list-f0-500.pdf.

\bibitem{klev}S. Klevansky, Rev. Mod. Phys. {\bf 64}, 649 (1992).

\bibitem{rob} 
R. D. Pisarski and F. Wilczek, Phys. Rev. D {\bf 29}, 338 (1984).










\bibitem{vac} 	
V. Skokov, B. Friman, E. Nakano, K. Redlich, and 
B.-J. Schaefer, 
Phys. Rev. D {\bf 82}, 034029 (2010).

\bibitem{lars}
R. Khan and L. T. Kyllingstad, 
AIP  Conf. Proc. {\bf 1343}, 504 (2011).
\bibitem{gupti}
U. S. Gupta, V. K. Tiwari,
Phys. Rev. D {\bf 85},  014010 (2012).

\bibitem{frg}
B.-J. Schaefer and J. Wambach, Nucl. Phys. A {\bf 757}, 479 (2005).


\bibitem{rebhan1}
R. Kobes, G. Kunstatter and A. Rebhan, Phys. Rev. Lett. {\bf 64}, 2992 (1990);
Nucl. Phys. B {\bf 355} 1 (1991).
\bibitem{rebhan2}
A. K. Rebhan, 
Phys. Rev. D {\bf 48}, R3967 (1993).
\bibitem{forth}P. Adhikari, J. O. Andersen, and Patrick Kneschke, (to be published).

 	

\end{thebibliography}
\bibliographystyle{apsrmp4-1}
\end{document}